\documentclass{epl}

\title{Motion of vortices implies chaos in Bohmian mechanics}

\author{D. A. Wisniacki\inst{1} \and F. E. R. Pujals\inst{2,3}}

\institute{
  \inst{1} Departamento de F\'\i sica ``J.J. Giambiagi'',
           FCEN, UBA, Pabell\'on 1, Ciudad Universitaria,
           1428 Buenos Aires, Argentina. \\
  \inst{2} Department of Mathematics, 
           University of Toronto,
           Toronto, Ontario, Canada M5S 3G3. \\
  \inst{3} IMPA-OS, 
           Dona Castorina 110, 22460-320 Rio de Janeiro, Brasil.
}

\newcommand{\te}{\theta}
\newcommand{\emp}{\emptyset}
\newcommand{\ga}{\gamma}

\begin{document}

\maketitle

\begin{abstract}
Bohmian mechanics is a causal interpretation of quantum mechanics
in which particles describe trajectories guided by the wave function. 
The dynamics in the vicinity of nodes of the 
wave function, usually called vortices, is regular if they are at rest. 
However, vortices generically move during time evolution of the system.     
We show that this movement is the origin of chaotic behavior of quantum 
trajectories.  
As an example, our general result is illustrated numerically  in the 
two-dimensional isotropic harmonic oscillator. 
\end{abstract}
De Broglie-Bohm's (BB) approach to quantum mechanics has experienced an increased
popularity in recent years.  This is due to the fact that it combines the 
accuracy of the standard quantum description with
the intuitive insight derived from the causal trajectory formalism, 
thus providing a powerful theoretical tool to understand the physical 
mechanisms underlying microscopic phenomena \cite{book,Durr}. 
Although the behavior of quantum trajectories is very different from
classical solutions it can be used to gain intuition in many physical phenomena.
Numerous examples can be found in different
areas of research. In particular, we
can mention studies of barrier tunneling in smooth potentials \cite{Wyatt},
the quantum back-reaction problem \cite{Prezhdo} and
ballistic transport of electrons in nanowires \cite{Wu1}.

According to the BB theory of quantum motion, a particle moves in a deterministic
orbit under the influence of the external potential and a quantum potential
generated by the wave function. This quantum potential can be very intricate
because it encodes information on wave interferences. Based on it, Bohm already 
predicted complex
behavior of the quantum trajectories in his seminal work \cite{Bohm1}. This was 
recently confirmed 
in several studies when presence of chaos
in various systems has been shown numerically \cite{bbchaos,Frisk,Wu2}. 
However, the mechanisms that cause such a complex behavior is still lacking.
In this letter we show that movement of the zeros of the wave function,
commonly known as vortices, implies chaos in the dynamics of quantum 
trajectories.
Such a movement perturbs the velocity field producing transverse homoclinic
orbits that generate the well known Smale horseshoes which is the origin of 
complex behavior. Our assertion is based on an analytical proof in a simplified model which 
resembled the velocity field near the vortices. In addition, we present a numerical 
study in a 2-D isotropic harmonic oscillator 
that displays a route to chaos dominated by this mechanism. It is
important to mention that there is no agreement in previous works about
the influence of vortices on the chaotic motion of quantum trajectories
\cite{Frisk,Wu2,Falsaperla,Valentini}.  
    
The fundamental equations in the BB theory are derived from the introduction
of the wave function in polar form,
$\psi(\textbf{r},t) \! = \! R(\textbf{r},t) \;
  {\rm e}^{{\rm i} S(\textbf{r},t)}$
(throughout the paper $\hbar$ is set equal to 1),
into the time--dependent Schr\"odinger equation, thus obtaining two
real equations:
%
\begin{eqnarray}
   & \displaystyle \frac{\partial R^2}{\partial t}+\nabla \! \cdot \!
   \displaystyle \left(R^2 \; \frac{\nabla S}{m} \right)=0, & \label{eq:1a}\\
   & \displaystyle \frac{\partial S}{\partial t}+\frac{(\nabla S)^2}{2m}
     +V-\displaystyle\frac{1}{2m} \frac{\nabla^2 R}{R} = 0, &
 \label{eq:1b}
\end{eqnarray}
which are the continuity and quantum Hamilton--Jacobi equations,
respectively.
The last term in the left--hand--side of Eq.~(\ref{eq:1b}) is the
so--called quantum potential, a non--local function determined by
the quantum state, which, together with $V$, determines the total
force acting on the system.
Similarly to what happens in the usual classical Hamilton--Jacobi theory,
quantum trajectories of a particle of mass $m$ can then be defined by 
means of the following velocity field equation:
%
\begin{equation}
   {\bf v}= \dot{\bf r} =\frac{1}{m} \nabla S=
\frac{i}{2 m}\frac{\psi {\bf \nabla} \psi^*-\psi^* {\bf \nabla} \psi}{|\psi|^2} .
 \label{eq:qt}
\end{equation}

Vortices appear naturally in the BB framework. They result from wave function 
interferences so they have no classical explanation. In systems without magnetic
field, the {\it bulk} vorticity ${\bf \nabla} \times {\bf v}$ in the probability
fluid is determined by the points where the phase $S$ is singular. This
may occur only at points where the wave function vanishes.
This condition is fulfilled by isolated points in a 2-D system and lines in 
a 3-D system.    
Due to the single-valuedness of the wave function, the circulation $\Gamma$ along any
closed contour $\xi$ encircling a vortex must be quantized, that is,
\begin{equation}
\Gamma=\int_{\xi} \dot{\bf r} d {\bf r}=\frac{2 \pi  n}{m},
\end{equation}
with $n$ an integer \cite{Dirac,Bialynicki1}. So, the velocity ${\bf v}$ must 
diverge as one approaches
a vortex. 
In fact,  the time dependent velocity field in the vicinity of
a vortex located at time $t$ in ${\bf r_v}(t)$ is given by
\begin{equation}
{\bf v}=\frac{-i}{2 m} \frac{\Big[{\bf r}-{\bf r_v}(t)\Big] \times {\bf w} \times {\bf w^*}}
                                 {\Big|\Big[{\bf r}-{\bf r_v}(t)\Big]{\bf .}{\bf w}\Big|^2},
\label{eq:vfield}
\end{equation}
where ${\bf w}\equiv {\bf \nabla} \psi({\bf r_v}(t))$ \cite{Falsaperla,Bialynicki2}.
We consider here 2-D systems but we belive that our results are valid 
for systems of higher dimensions.


Before presenting our analytical results we show a numerical simulation of quantum trajectories
in a system consisting of a particle of unit mass in a 2-D  isotropic harmonic oscillator. 
We have set the angular frequency $\omega=1$, so the Hamiltonian 
of the system results
\begin{equation}
H= -\frac{1}{2} (\frac{\partial^2}{\partial x^2}+\frac{\partial^2}{\partial y^2})
+\frac{1}{2}(x^{2}+y^{2}).
\label{eq:hami}
\end{equation}
The eigenenergies are 
$E_{n_x n_y} =  n_{x}  +  n_{y}  +1$ and the eigenfunctions  
$\phi_{n_x n_y}(x,y)  =  \exp (-\frac{1}{2}(x^{2}+y^{2}))
H_{n_x}(x) H_{n_y}(y)/\sqrt{\pi \;2^{n_{x}+n_{y}}\; n_x!\; n_y!}$ 
with $n_x=0,1,...$, $n_y=0,1,...$ where $H_n$ is the n-th degree Hermite polynomial. 

We have chosen the following general combination
of the first three eigenstates of the Hamiltonian of Eq.~(\ref{eq:hami}) as initial state 
\begin{equation} 
\psi_0 =a \phi_{0 0} + b \exp(-i \gamma_1) \phi_{1 0}+c \exp(-i \gamma_2) \phi_{0 1},
\label{eq:psi}
\end{equation}
with $a,b,c,\gamma_1$ and $\gamma_2$ real numbers and $a^2+b^2+c^2=1$ (the normalization condition). 
A remarkable point is that this state generates a periodic time 
dependent velocity field with only one vortex. Moreover, the trajectory of the vortex 
can be obtained analytically resulting
\begin{equation} 
{\bf r_v}(t)=(x_v(t),y_v(t))=(\frac{a}{\sqrt{2} b}  \frac{\sin(\gamma_2-t)}{\sin(\gamma_1-\gamma_2)},
\frac{a}{\sqrt{2} c}\frac{\sin(\gamma_1-t)}{\sin(\gamma_1-\gamma_2)}).
\label{eq:mov}
\end{equation}
This fact allows us to see the influence
of the movement of a vortex in the dynamics of the quantum trajectories, without taking into account 
the possibility of instantaneous creation or annihilation of a vortex pair with opposite
circulation \cite{Hirshfelder,Bialynicki2}. This important phenomenon will be studied elsewhere \cite{Wisniacki-f}. 

\begin{figure}
\twofigures[scale=0.44]{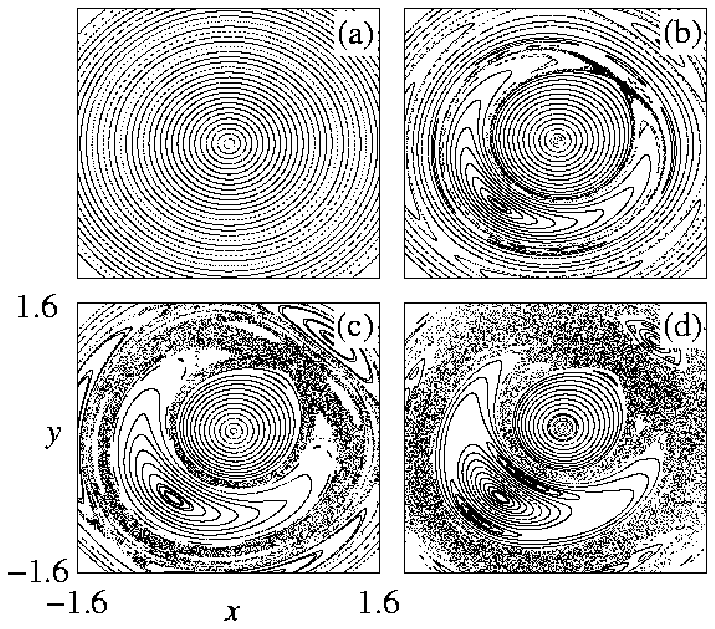}{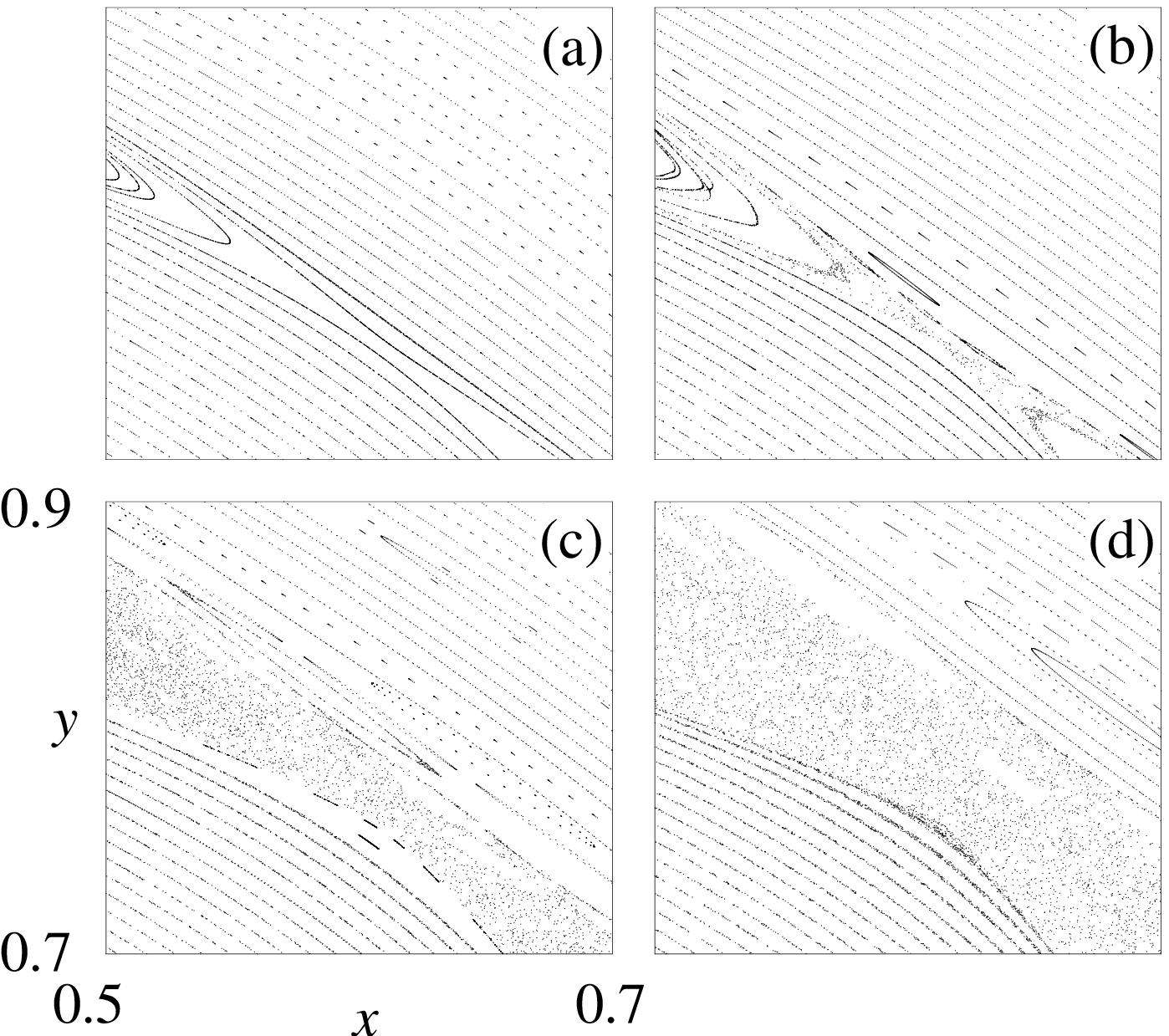}

\caption{Poincare surface of section for the quantum trajectories generated by the
wave function of Eq.~(\ref{eq:psi}) with $b=c$ and $a/b=0$ (a) , $0.0553$ (b), 0.1138 
(c) and 0.17651 (d) with fixed value of $\gamma_1=3.876968$ and $\gamma_2= 2.684916$. 
The trajectories
of the vortex of the corresponding wave functions are shown in Fig. ~\ref{fig:3}.}
\label{fig:1}
\caption{Part of the Poincare surface of section for the quantum trajectories generated by the
wave function of Eq.~(\ref{eq:psi}) with small values of $a/b$. The parameter $b=c$ and 
$a/b=0.01082$ (a) , 0.02175 (b), 0.0328 (c) and 0.0440 (d) with fixed value of 
$\gamma_1=3.876968$ and $\gamma_2= 2.684916$.}
\label{fig:2}
\end{figure}

\begin{figure}[h]
 \onefigure[scale=0.28]{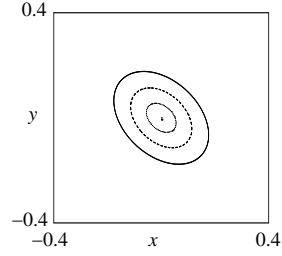}
 \caption{Path described by the vortex [Eq.~(\ref{eq:mov})] of the velocity field  
generated by wave functions of Eq.~(\ref{eq:psi})  with $b=c$ and $a/b=0$ (filled circle) , 
$0.0553$ (dotted line), 0.1138 (dashed line) and 0.17651 (solid line) with fixed value of 
$\gamma_1=3.876968$ and $\gamma_2= 2.684916$.}
\label{fig:3}
\end{figure}

The non-autonomous velocity field generated by the wave function of 
Eq.~(\ref{eq:psi}) is periodic so the best surface of section is given 
by fixing  $t=2 \pi n$ with $n=0,1,....$ (also called 
a stroboscopic view). Fig.~\ref{fig:1} shows   
surfaces of section with $b=c$ and $a/b=0, 0.0553, 0.1138$ and $0.17651$ with
fixed value of $\gamma_1=3.876968$ and $\gamma_2= 2.684916$. 
The trajectories of
the vortex for the cases studied in Fig.~\ref{fig:1} are
plotted in  Fig.~\ref{fig:3}. A clear  transition to chaos appears
as the parameter $a/b$ is increased. If the position of vortex is fixed
the trajectories are regular and no chaos is
present [see Fig.~\ref{fig:1}(a)] . However, irregular dynamics is observed 
for small $a/b$ [Fig.~\ref{fig:1}(b)].
The transition to irregular dynamics is shown in Fig.~\ref{fig:2}.   
The movement of the vortex produces a saddle point near $(0.6,0.75)$ and
their stable and unstable manifolds have a topological transverse intersection generating the
well known homoclinic tangle \cite{P}. 

Now we will show analytical results that explain the numerical experiments
presented before. Our starting point is the following model:
a particle
of unit mass on the plane in the velocity field of Eq.~(\ref{eq:vfield}) with the
constraints that the trajectory of the vortex is a time periodic curve
and $w_x=i w_y$ \cite{assu}.
Thus, the non-autonomous vector field  is equal to

\begin{eqnarray}
v_x &=& \frac{-(y-y_v(t))}{(x-x_v(t))^2+(y-y_v(t))^2} \nonumber \\
v_y &=& \frac{(x-x_v(t))}{(x-x_v(t))^2+(y-y_y(t))^2}\;.
\label{eq:vfield2}
\end{eqnarray}

Taking $\bar{x}= x-x_v(t)\, \bar{y}=y-y_v(t)$ and writing in polar coordinates
($\bar{x}= r \cos (\te) , \, \bar{y}=r \sin(\te)$), the velocity field of 
Eq.~(\ref{eq:vfield2}) results

\begin{eqnarray*}
v_r &=& r [\sin(\te) y_v(t)+x_v(t) \cos(\te)]\\
v_{\te} &=& \frac{1}{r}+ \cos(\te) y_v(t)-x_v(t)\sin(\te).
\end{eqnarray*}

This non-autonomous velocity field can be seen as a perturbation of the autonomous 
velocity field 
${\bf v}_0 \equiv (0,\frac{1}{r})$, with 
${\bf G}(r,\te,t)\equiv (r [\sin(\te) y_v(t)+x_v(t) \cos(\te)],
\cos(\te) y_v(t)-x_v(t)\sin(\te))$ the time-periodic perturbation. Note that the field is induced 
by the time dependent Hamiltonian  

\begin{equation}
H(r,\te,t)= \frac{1}{2}\log(r)+r[\cos(\te) y_v(t)-\sin(\te) x_v(t)].
\label{hami}
\end{equation}
We consider periodic curves ${\bf r_v}(t)$ such that

\begin{equation}\label{cond}
\int_0^{T_0} \cos(\te)y_v(t)-x_v(t) \sin(\te) dtds\neq 0.
\end{equation}

Under these hypothesis the following property can be proved: 

\vskip  3pt

{\it There exists a saddle periodic 
orbit of the flow associated to the vector field of Eq.~(\ref{eq:vfield2}), 
exhibiting a homoclinic transversal intersection.}  

\vskip  3pt

This result, which implies that quantum trajectories show topological chaos,
is the main finding of our work . 
We illustrate here the geometrical arguments of the proof and 
we leave the full details for a future publication \cite{Pujals}.
We notice that a periodic orbit of saddle type exhibits a homoclinic 
transversal intersection if their stable and unstable manifolds intersect 
each other 
and the tangents of the manifolds are not colinear at the intersection.
Transversal homoclinic intersections (or homoclinic
tangles) beat at the heart of chaos. This is because in the
region of a homoclinic tangle, initial conditions are
subject to a violent stretching and folding process,
the two essential ingredients for chaotic behavior. 

\begin{figure}[t]
\onefigure[scale=0.55]{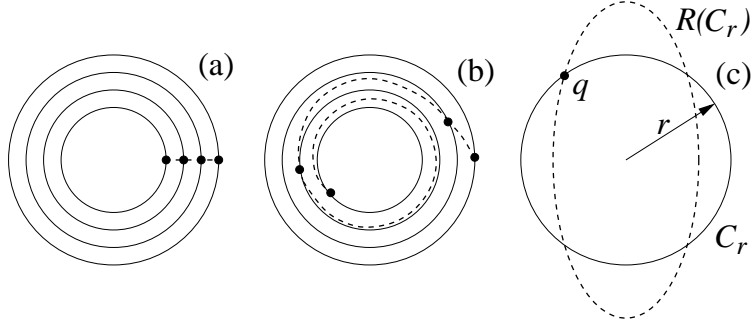}
 \caption{(a) Schematic plot of the invariants of the non-perturbed map $R^0$ generated
by the autonomous velocity field ${\bf v}_0$ (solid lines). With dashed line it is plotted a segment 
with $\te=0$.  
(b) Non-perturbed mapped $R^0$ of the segment with $\te=0$ (dashed line).
(c) It is showed with dashed line the image of the perturbed map $R$ of a circle 
$C_r$ with radius $r$ (solid line). A point $q\in R(C_r)\cap C_r$ is also plotted.}
\label{fig:4}
\end{figure}

Let us start to show the main result of the letter
considering some important charateristics and properties of
the flows generated by the velocity fields ${\bf v}_0$ and 
${\bf v}$. 
The flow  $\Phi^0_t$ associated with the autonomous velocity 
field ${\bf v}_0$  is defined for every $(\bar{x},\bar{y})\neq (0,0)$. 
Thus, given a positive time $T_0$, the map $R^0= \Phi^0_{T_0}:
\Re^2\setminus\{(0,0)\}\to \Re^2\setminus\{(0,0)\}$ is well defined.
It is straightforward to show that the map $R^0$ written in polar 
coordinates results $R^0=(R^0_r(r,\te),R^0_\te(r,\te))=(r,\te+\frac{1}{r}T_0)$.
Map $R^0$ keeps
invariant the set of points with same radius; i.e., it keeps
invariant the  circles. 
Moreover, the circles are rotated with a
rate of rotation inversely proportional  to the radius [see Fig.~\ref{fig:4}(a) and (b)]. 
Observe that map $R^0$ resembles a twist map with the difference
that in the present case, the rate of rotation grows to infinity when the 
radius is reduced. In this respect, it is important to mention that for generic conservative 
perturbations of the twist map the existence of homoclinic points associated to a saddle periodic
point was proved \cite{Z} . This result was also extended to time periodic perturbations of a flow
which exhibit an elliptic singularity \cite{GH}.

We have assumed that the vortex moves periodically along
a curve, that is,  ${\bf r_v}(t)={\bf r_v}(t+T_0)$. 
Then, the time dependent 
velocity field
${\bf v}$ induces a flow $\Phi_t$ which is defined for every
$(\bar{x},\bar{y})\neq \{(0,0)\}$. 
So, the map $R= \Phi_{T_0}: \Re^2\setminus\{(0,0)\}\to \Re^2\setminus\{(0,0)\}$ is well defined. 
Note that flow $\Phi_t$ is generated by a time periodic Hamiltonian [Eq.~(\ref{hami})], 
so $R$ is a conservative map. Also, $R$ can be extended
continuously to (0,0) defining $R(0,0)\equiv(0,0)$ [note that
$R \rightarrow 0$ when $(\bar{x},\bar{y}) \rightarrow (0,0)$].
Then, it follows that map $R$ verifies:

{\it Property A:} Like $R^0$, map $R$ also has the property that circles are
rotated with a rate of rotation inversely proportional to the
radius [see Fig.~\ref{fig:4}(a) and (b)]. In other words, if map $R$ is written in polar
coordinates $R(r,\te)=(R_r(r,\te),R_\te(r,\te))$,
then $\partial_r R_\te$ is of the  order of $1/{r^2}$. 

{\it Property B:} The image of a small circle of radio $r$
intersects transversally this circle; i.e, $R(C_r)\cap C_r\neq \emp$ and  
if $q\in R(C_r)\cap C_r$ then the tangent
to $R(C_r)$ and to $C_r$ in $q$ are not collinear [see Fig.~\ref{fig:4}(c)].

From {\it properties A} and {\it B} of the perturbed map $R$ it follows that for
arbitrarily small $r$ the map $R$ has a fixed point $p_0$ with
radius smaller than $r$ exhibiting homoclinic transversal
point. Of course, this result implies that the vector field of 
Eq.~(\ref{eq:vfield2}) has a saddle periodic orbit with an homoclinic 
transversal intersection. 

\begin{figure}
\onefigure[scale=0.45]{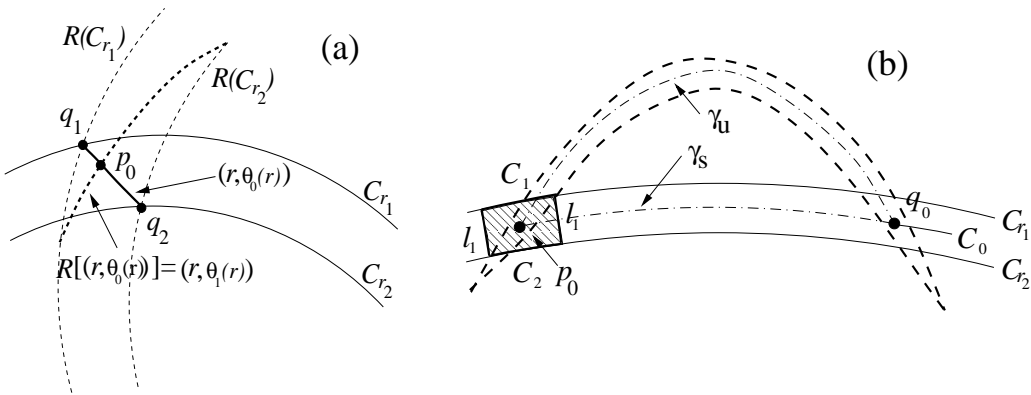}
\caption{Schematic plots to show  
the existence of a saddle periodic point $p_0$ of the map $R$ with a homoclinic 
transversal intersection. (a) The curve 
$r\to (r,\te_0(r))$ is plotted with thick solid line, 
and the curve
$r\to R((r,\te_0(r)))=(r,\te_1(r))$ is plotted with thick dashed line.
Two invariants circles
($C_{r_1}$ and $C_{r_2}$) and their perturbed mapping are also plotted. Note the transversal
intersection between the circles an their respective mapping (points $q_1$ and $q_2$).
(b) The rectangle $B$ (shaded area) is bounded by two pieces of arcs $C_1$ and $C_2$
contained in $C_{r_1}$ and $C_{r_2}$ (where $r_1$ and $r_2$ are closed to $r_0$ verifying 
$r_1<r_0<r_2$) and two segments
$l_1$ and $l_2$ contained in two different rays of constant angle. 
The segments $\ga_u$ and $\ga_s$, that connects the periodic point $p_0$ with the 
transversal intersection $q_0$, are ploted with dash-dotted lines.  The mapping $R(B)$ of the considered rectangle is also ploted with thick dashed line.}
\label{fig:5}
\end{figure}

We show the previous result in two steps. First, let us display 
that the perturbed map $R$ has a fixed point. {\it Property B} guarantees the
existence of the curve $(r,\te_0(r))$ plotted in Fig.~\ref{fig:5}(a).
Note that $\te_0(r)$ is the angular coordinate of a point on the 
intersection of a circle  $C_r$ with its image $R(C_r)$. 
A point on such a curve is mapped to other point with the same radius $r$
but with a different angle $\te_1(r)$, 
\[
R(r,\te_0(r))=(R_r(r,\te_0(r)),R_\te(r,\te_0(r))=(r,\te_1(r)).
\]
It is clear that map $R$ has a fixed point if there exist an $r_0$ such
that $\te_0(r_0)=\te_1(r_0)$. In other words,  
curves  $(r,\te_0(r))$ and $(r,\te_1(r))$ of 
Fig.~\ref{fig:5}(a) intersect at $r_0$.   
From {\it property A} follows
that the variation of $\te_1(r)$ is larger than the variation of
$\te_0(r)$; in fact, the derivative of $\te_1(r)$ is of the order
of $1/{r^2}$. This fact guarantees the existence of $r_0$ in the vicinity
of $r \to 0$ \cite{Pujals}.

Now we will see that the mentioned fixed point has a 
transversal homoclinic intersection of their stable and unstable manifolds. 
We recall that the stable manifold
is the set of points that converges to the fixed point by
forward iteration of the dynamic. Conversely, 
the points on the unstable manifold converge to the fixed point by
backward iteration.
In Fig. ~\ref{fig:5}(b) we consider a circle $C_{0}$ of radius $r_0$ containing the fixed point
$p_0$. Due to {\it property B}, $C_0$ and its image $R(C_0)$ has al least an
additional point of intersection denoted by $q_0$. Points $p_0$ and $q_0$ 
are connected by the segments $\ga_s$ and $\ga_u$ of $C_0$ and $R(C_0)$
respectively.
Let us consider the shaded rectangle $B$ of Fig.~\ref{fig:5}(b) that contains 
the fixed point $p_0$.
Using {\it property A}, we have deduced that segments of constant angle $l_1$ and $l_2$  
are stretched by $R$, and this shows that rectangle $B$ 
is contracted by $R$ along directions close
to the tangent of the circles $C_r$ and expanded along
vectors close to the tangent of $R(C_r)$.
This is displayed in Fig.~\ref{fig:5}(b) where the mapping of the rectangle $B$ is plotted 
with thick dashed line. 
Moreover, we have proved in Ref. \cite{Pujals} that $R(B)$ is close to $\ga_u$ 
and intersects $C_0$
in a point near $q_0$, and that, $R^{-1}(B)$ is close to $\ga_s$ and
intersects $C_0$ near $q_0$. 
This implies that the unstable manifold of $p_0$ is close to $\ga_u$ and the stable 
manifold is
close to $\ga_s$. We recall that $\ga_u$ cross $\ga_u$ at $q_0$, then
the stable and unstable manifolds of $p_0$ have a transversal intersection near $q_0$. 

In summary, we have found an universal mechanism leading to quantum trajectories 
having chaotic behavior. We have shown that the movement of
vortices is a generic time dependent perturbation of an autonomous
velocity field which creates a saddle periodic orbit with a transversal
homoclinic intersection of their stable and unstable manifolds. 
This transversal intersection 
generates the well known Smale horseshoe which is the origin of complexity . 
Our results should be useful due to the fact that such deterministic quantum orbits
are an important theoretical tool for understanding and interpreting several 
processes in different fields. On the other hand, our geometrical analysis
of a singular velocity field could be important for both theoretical and applied problems of 
dynamical systems, as for example advection in non-stationary fluids \cite{fluids}.

\begin{acknowledgments}

This work received financial support from CONICET and UBACYT (X248). We thank G. Lozano and 
F. Duffy for useful comments. 
\end{acknowledgments}
%


\begin{thebibliography}{0}
\bibitem{assu} 
This is the simplest toy model because $w_x$ and $w_y$ are time dependent functions given by the derivatives of the wave function.
\end{thebibliography}
\end{document}